# Integer Lattice Gases at Equilibrium

**Alex Khaneles**
Vinshtok Software Development, LLC
19 Kingsland Circle, Monmouth Junction NJ 08852, USA
e-mail: akhaneles@yahoo.com

**ABSTRACT**

*Integer lattice gas automata can be utilized as building blocks in statistical mechanics. The presented deterministic and reversible automaton generates semiclassical statistical distributions. A possible approach to Bose-Einstein statistics from cellular automata position is discussed.*

**Keywords**: cellular automata, statistical physics, quantum statistics

**AMS Subject Classification**: 82B20, 82B30, 82-04, 82B05, 82B10

## 1. INTRODUCTION

A computer technique of Cellular Automata (CA) was invented by Stanislaw Ulam and John von Neumann in the late 1940's. This method is capable of simulating a system consisting of multiple identical elements and has found a variety of scientific applications including modeling in physics. One class of cellular models – lattice gases – was especially successful in the areas of hydrodynamics and turbulence. These models use point particles moving on a discrete grid and imitate molecular dynamics.

In such modeling, a cellular grid is often considered to be a simplified discrete substitution to the space continuum with finite spatial resolution. The local, deterministic and reversible character of some CA rules appears appropriate to express microscopic dynamics. At the same time, the discrete grid has certain limitations as a model of space. Although the lack of isotropy was overcome in hydrodynamics applications by a triangular lattice in the FHP model [6], deficiency of rotational symmetry is still apparent at the fundamental level [8]. On the other hand, clear-cut locality is not suitable in quantum physics (with Bell's inequality).

Another area of feasible CA applications is statistical mechanics, which operates with an ensemble of states rather than with space and where questions in interpretation of space are secondary. CA can simulate such an ensemble as a set of cells. An attractive feature of the method is the ability to keep track of each and all of cellular elements during evolution. It is possible to perform any measurements or observations without disturbing the system. Although full knowledge of the system requires intense calculations, today's personal computer is capable of handling about $10^6$ cells and simple models in reasonable time. This computational power is adequate for analysis of thermodynamic behavior of some important model systems.

It is possible to express the ideas of microscopic dynamics as CA rule. Subsequently one can observe the manifestation of this microdynamics in a statistical embodiment as the evolution occurs. Thus, the technique simplifies study of macroscopic behavior and may provide new insights, for example, to questions like: how can irreversible macroscopic behavior arise from time symmetric microscopic dynamics?

One of the advanced lattice gas methods is an integer lattice gas that generalizes this kind of automata to include arbitrary numbers of particles moving in each lattice direction. It was introduced first on a triangular grid for hydrodynamic modeling as an expansion of FHP gas [1]. Integer cell characteristics enrich behavior of cellular system and allow generating additional statistical distributions.

In this paper we utilize an integer lattice gas of massless particles implemented on a rectangular lattice. The statistical behavior of the automaton shows a similarity to Planck's harmonic oscillators and reproduces certain aspects of semi-classical statistics at equilibrium. Depending on the CA rule, the resulting statistical distributions can be shaped differently. We will discuss the possibility of obtaining Bose-Einstein statistics from a deterministic CA position.

## 2. INTEGER HPP GAS

HPP gas (named after its inventors) [7] is a discrete abstraction of an ideal gas for CA modeling where one bit particles are moving along four possible directions on the two-dimensional rectangular lattice. Its behavior was investigated in detail, and illustrations of sound wave propagation, some optical phenomena and more [13, 4] are available. Presented here, the automaton is an extension of HPP gas which accommodates any number of particles in each cell moving in each lattice direction. We will call this automaton Integer HPP gas (IHPP).

(Notions that are specific to the context of this paper are printed in italics when used for the first time.)

### The Lattice

We utilize a two-dimensional rectangular CA grid where each cell has four neighbors (the von Neumann neighborhood). The lattice has finite size, and the boundaries are connected to opposite sides (torus-like topology).

### The Cell State

The cell state consists of four integer characteristics: ${}^{+}E_x$; ${}^{-}E_x$; ${}^{+}E_y$; ${}^{-}E_y$. Each integer contains the number of particles moving in one of four possible directions on the grid (${}^{+}x$, ${}^{-}x$, ${}^{+}y$, and ${}^{-}y$).

### The Rule

The particles are moving along axes without changing direction until they meet others moving in the opposite direction (a head-on *collision* or *scattering*) that results in changing their direction on the *perpendicular*.

The rule forms the next state of neighboring cells from the current state of a single cell and should be performed for each cell in the grid in each evolution step:

1. Calculate the numbers of collided/scattered particles along $x$ and $y$ as:
   (the numbers of $x$ particles scattered into $y$ ) $= \min({}^{+}E_x;{}^{-}E_x)$
   and
   (the numbers of $y$ particles scattered into $x$ ) $= \min({}^{+}E_y;{}^{-}E_y)$;

2. All particles are moving from the current cell to the neighboring cells. The scattered particles travel in new directions. The remaining particles travel in unchanged directions.

The rule articulates the ideas of continuation of motion without interactions and detailed balancing in interactions. It is simpler than the HPP rule in that there is no necessity to exclude scattering into the cell characteristics which are already occupied.

## 3. IHPP GAS FEATURES

Since the interest of this paper is IHPP gas in a state of equilibrium, we mention here some related features that are inherited from HPP gas.

IHPP gas conserves *energy* - the total number of particles in the system:

$$(energy) = \sum_N ({}^{+}E_x + {}^{-}E_x + {}^{+}E_y + {}^{-}E_y) = const ,$$

where *N* is a number of cells in the system.

It conserves *momentum components* – the differences of the total numbers of particles moving in opposite directions:

$$(\textit{momentum along } x) = \sum_N ({}^{+}E_x - {}^{-}E_x) = const\,;$$

$$(\textit{momentum along } y) = \sum_N ({}^{+}E_y - {}^{-}E_y) = const\,.$$

Growing disorder in the cellular system (from certain initial states) to a state of equilibrium illustrates the Second Law of Thermodynamics. No information is lost during evolutions, and at any moment we can reverse the direction of time and bring the system back step by step to its initial state. It demonstrates how seemingly irreversible behavior in kinetic theory could arise from fully reversible underlying dynamics ('reversibility paradox' of Josef Loschmidt, 1876) and Boltzmann's understanding of the statistical nature of the Second Law. Illustrations of this kind with HPP gas have been presented in literature previously [13, 4, 14][1].

Using numbers instead of bits as characteristics of the cell state in our IHPP gas leads to an extra dimension in the behavior of the system. Each integer characteristic in the cell state can be regarded as an occupation number ($n$). To analyze this extra dimension we have an additional instrument - statistical distributions.

**Initialization**

The attributes of cells can be initialized uniformly with the same numbers or with different numbers. If *the symmetry is broken* at initialization then such a system will keep disorder growing during evolutions and will come to the state of stable equilibrium. The more irregularities that are created at initialization, the faster disorder will grow. In our simulation $\sim 3 \cdot 10^5$ evolutions are considered as sufficient *relaxation time* for the grid of $N \sim 2 \cdot 10^6$ cells.

Some specifics of initialization should be mentioned here. The smallest number used in initialization will stay as a minimum occupation number over all evolutions. It will be named *zero point energy*. It is possible to set a step in occupation numbers as well. If we use only numbers (2; 5; 8; 11) with particular interval between them (in this sample - 3) to initialize the cell states, the spectrum will maintain the same step size (2; 5; 8; 11; 14; 17; …) throughout the evolutions.

To summarize, the initialization sets three parameters which will affect the statistical picture at equilibrium:

1. Zero point energy: $z$ - any (including negative) integer number;

2. The mean occupation number: $\overline{n}$ ($\overline{n} > z$);

3. The step in occupation numbers: $s$ - any positive integer number.

These parameters will be accommodated to analyze the results of simulation as well.

**Occupation Number Distribution**

Figure 1 shows the occupation number distributions for three simulation series. The exponential shape of these distributions corresponds to standard expectations of statistical mechanics. We compare the simulation results with the empirical function:

$$f(n) = N \cdot a \cdot s \cdot \frac{(\overline{n} - z)^{n-z}}{(\overline{n} - z + 1)^{n-z+1}} = N \cdot a \cdot s \cdot \frac{1}{\overline{n} - z + 1}\left(1 + \frac{1}{\overline{n} - z}\right)^{-(n-z)} \quad (1)$$

where $a$ is the number of attributes in each cell ($a$ = 4 for the two-dimensional system).

If we limit simulation conditions to $z$ = 0 and $s$ = 1, the number of parameters in the generic formula (1) will be reduced and the formula itself will come to what is familiar for Planck's oscillators (a photon number distribution for a single mode of thermal radiation):

$$\frac{f(n)}{N \cdot a} = \frac{\bar{n}^n}{(\bar{n}+1)^{n+1}} = \frac{1}{\bar{n}+1}\left(1+\frac{1}{\bar{n}}\right)^{-n} . \quad (2)$$

For these exponential distributions (2) it would be natural to introduce such notions as *quanta of energy* ($\varepsilon_0$) and *temperature* ($T$) and to use Planck's distribution function to describe our IHPP gas:

$$1+\frac{1}{\bar{n}} = e^{\varepsilon_0/k_B T} , \quad (3)$$

where $k_B$ has the same meaning as Boltzmann constant, or

$$\bar{n} = \frac{1}{e^{\varepsilon_0/k_B T}-1} . \quad (4)$$

The logic of our model is purely discrete, and real numbers are used only to compare the results of simulation with commonly used mean values, formulas, etc.

IHPP gas illustrates Planck's harmonic oscillators (only a single mode of thermal radiation). Planck introduced discontinuity of energy to explain blackbody radiation. Therefore IHPP gas can illustrate other ideas historically related to the explanation of thermal radiation and to obtain Wien's distribution (see Appendix A and Figure 2).

It has been known that Boltzmann statistics would bring us to Wien's formula and not to Planck's radiation law [2]. The next question was *'what changes must be made in the classical statistics in order that it may become possible to deduce Planck's radiation law by purely statistical reasoning, without making use of the roundabout road by way of an absorbing and emitting oscillator'* [2]. Satyendranath Bose answered this question with the invention of quantum statistics in 1924 [3].

What would it mean in terms of CA to get Planck's radiation law without '*the roundabout road*'? In our picture, the roots of statistics are in the CA rule and to change statistics means to change the rule. Could Bose-Einstein statistics be reproduced this way?

## 4. DETERMINISTIC CA OUTLOOK ON BOSE-EINSTEIN STATISTICS

> *'Bose's arguments divert Planck's law of all supererogatory elements of electromagnetic theory and base its derivation on the bare essentials. … I believe there had been no such successful shot in the dark since Planck introduced the quantum in 1900.'*
>
> A. Pais [11]

To come to the blackbody energy spectrum, Bose presumed that Planck's quanta form a single mode of thermal radiation and introduced multiple modes and a recipe to count the densities of different modes. We have seen how statistical behavior of Planck's quanta can arise from deterministic microscopic dynamics of IHPP gas. Could we capitalize on the same basis (as Bose did) to obtain quantum statistics and blackbody radiation distribution from CA rule?

The multiple modes (species) of quanta in Bose reasoning are characterized by their energy $\varepsilon_s$ ($s$ = 0 to $s = \infty$). Our set of cells would represent only the single mode in an infinite number of modes. Although to make the transition from Planck distribution (4) to Bose-Einstein distribution:

$$\bar{n}_s = \frac{1}{e^{\varepsilon_s / k_B T} - 1} \ , \tag{5}$$

only the index $s$ has to be added.

While incorporating these ideas in a formula is straightforward, their application in CA logic is not apparent. Does this index mean that we should multiply entities of cell states and put them into an infinite number of layers (modes)? It would be a vast increase in the amount of information. It is not clear how to manage this information and redistribute energy between modes. Simply adding another dimension to the cellular system is not appropriate – the number of cells (density of states) in each mode should be a function of the energy $\varepsilon_s$.

Should we try to fulfill such significant changes in the entire structure or are there other more attractive directions for further development? The indistinguishable states and multiple modes of radiation not only break the rules of Boltzmann statistics (what Bose did not suspect in 1924 [11]), they also do not fit well into the concept of deterministic CA. As long as it was fairly easy to arrive at Planck's oscillators with IHPP gas and to obtain the shape of the Wien's law (Appendix A), it should not be so complicated to approach the seemingly not much different Planck's radiation law. An effort to find an alternative approach with CA would be supported at least by Occam's razor: entities should not be multiplied unnecessarily. With lattice gas technique we can increase complexity of the rule and the amount of information gradually.

As demonstrated, IHPP gas described here generates classical statistical behavior and replicates Planck's oscillators but has not much to do with quantum statistics. Nonclassical features could be introduced by including a new content into the integer lattice gas rule and as a result altering the statistical distributions[2]. Perhaps we just have to deploy additional vital physical principles in CA to exhibit the behavior that is associated with photons and Bose-Einstein statistics.

In the evolution of the explanation of blackbody radiation, attention has been gradually deflected from its electromagnetic nature (to *'the bare essentials'*). With CA we can try to move in the opposite direction – not to divert *'supererogatory elements'* but to take them in as essentials and prepare the rule of microscopic dynamics from the elements of Maxwell's electrodynamics to study statistical consequences.

## 5. PROSPECTIVE DEVELOPMENT

The first step in this new trend would be to employ two integer cell state attributes in each lattice direction (instead of one in IHPP gas). These attributes could be interpreted as components of energy or electrical and magnetic fields. Using two integer attributes in each direction would provide a sufficient amount of information to introduce elements of light polarization or two possible photon spin orientations. In order to develop new ideas in the CA rule, however, it is important to utilize different grids such as triangular or 3-dimensional. More than two axes in the grid would allow for different scattering directions ($x \to y \to z \to \ldots$, $x \to z \to y \to \ldots$). Allowing a conversion of energy between two components with different scattering directions can bring new features into the basic statistical picture. This kind of CA could become a useful instrument in furthering our understanding of quantum statistics and its connection to spin. The nature of this connection has been considered to be quite complicated [12, 10, 5].

## 6. CONCLUSION

The integer lattice gas described here generates semiclassical statistical distributions from deterministic and reversible CA rule based on the ideas of continuation of motion without interactions and detailed balancing in interactions. The lattice gases of this kind do allow building different statistics and might be helpful in understanding the origins of quantum statistics. To produce new statistics, the CA rule can be expanded or changed to accommodate other (not mechanical) physical concepts. Future prospects of this work are to consider two cell characteristics in each direction and investigate possible relationship between them. Such CA rules could resemble ideas of Maxwell electrodynamics with electric and magnetic field components.

## Acknowledgements

The author would like to thank Valeri Golovlev for comprehensive discussions and valuable suggestions and Catherine Smith for critical comments.

## Endnotes

[1] IHPP rule is not the best to demonstrate reversibility. A rule of this kind can be time-symmetrical i.e. remain exactly the same for both time directions. IHPP gas was chosen here because its predecessor (HPP) has been well-investigated and some attractive physical interpretations have been provided already.

[2] The changes in the distributions can be shown even without any significant novelties in the CA rule. For example, two or three identical IHPP cellular systems can be thought of as one with doubled or tripled number of cell characteristics in each direction and the same rule for each subset. Then the occupation number or energy in one direction would be a sum of two or three components and corresponding distributions obviously would be very different from original distribution for one-component variable (see Figure 2 as a sample for three-component variables).

## APPENDIX A: THREE-DIMENSIONAL IHPP GAS AND WIEN'S DISTRIBUTION

Lord Rayleigh interpreted blackbody radiation as rest waves in a three-dimensional cavity (the detailed history of this development at the beginning of 20th century can be found in [9]). The permissible modes of electromagnetic field vibrations were governed by three integers $k_1, k_2, k_3$:

$\lambda = 2L \big/ \sqrt{k_1^2 + k_2^2 + k_3^2}$ , where $\lambda$ is a wavelength, $L$ – the side of a cubical cavity.

The integer numbers – $k_1, k_2, k_3$ – can be thought of as discrete momentum (or electric field) components. This approach led to "ultra-violet catastrophe".

IHPP gas replicates Planck's oscillators and is able to illustrate a mixture of Planck's and Rayleigh's ideas. We could construct triplets like ($k_1, k_2, k_3$) from cell state components, if our cellular system were three-dimensional. Expansion into three dimensions can be done quite easily. The 3d cell state will contain six (instead of four) integer characteristics: ${}^{+}E_x$; ${}^{-}E_x$; ${}^{+}E_y$; ${}^{-}E_y$, ${}^{+}E_z$, and ${}^{-}E_z$, corresponding to six possible directions on the 3d grid.

The 3d IHPP rule will not be much different from the 2d one either. In the 2d rule, the scattering was going from $x$ direction into $y$ and back from $y$ into $x$. In the 3d, it could be arranged from $x$ into $y$; from $y$ into $z$; and from $z$ into $x$. This rule will exhibit the same kind of statistical behavior that we have seen already for 2d lattice gas. The occupation number distribution (Figure 1) for the 3d cellular system will look exactly the same.

We can imagine that the components of one cell can act together and define composite variables to picture this idea:

$$^{+}n = {}^{+}E_x + {}^{+}E_y + {}^{+}E_z\,; \qquad {}^{-}n = {}^{-}E_x + {}^{-}E_y + {}^{-}E_z\,; \tag{A.1}$$

$$^{+}n_m = \sqrt{{}^{+}E_x^2 + {}^{+}E_y^2 + {}^{+}E_z^2}\,; \qquad {}^{-}n_m = \sqrt{{}^{-}E_x^2 + {}^{-}E_y^2 + {}^{-}E_z^2}\,. \tag{A.2}$$

It does not matter how to divide the cell components into groups of three, as long as the IHPP rule does not generate correlations between them.

Figure 2 presents distribution of triplets by ${}^{\pm}n$ (${}^{+}n$ and ${}^{-}n$) and another one of the same kind by ${}^{\pm}n_m$ (${}^{+}n_m$ and ${}^{-}n_m$). The distributions can be described by function:

$$f(n) = \frac{N \cdot s}{(\overline{n} - z)^3}(n - 3z)^2\left(1 + \frac{1}{\overline{n} - z}\right)^{-(n-3z)} \tag{A.3}$$

The function parameters ($\overline{n}$ , $z$ , $s$ ) for the first distribution (by ${}^{\pm}n$ ) are the same as those used for simulation. The second distribution (by ${}^{\pm}n_m$ ) has similar shape but the value of $\overline{n}$ is much less.

If we accept $z$ = 0, $s$ = 1 and apply Planck distribution (3), this function would come to:

$$f(n) = \frac{N}{\overline{n}^3}n^2\left(1 + \frac{1}{\overline{n}}\right)^{-n} = \frac{N}{\overline{n}^3}n^2 e^{-n\varepsilon_0/k_B T} \tag{A.4}$$

Thus, the straightforward combination of elements of Planck's and Rayleigh's ideas bring us to the shape of Wien's distribution (A.4). To get energy distribution it should be multiplied by $n\varepsilon_0$. We do not pay attention here to the geometrical factor before $n^2 e^{-n\varepsilon_0/k_B T}$ in the formula since distance, speed of light, and volume is not defined in our model.

Both composite variables (A.1) and (A.2) are somewhat artificial, as far as the CA rule does not lead to correlations between cell state components. Even less justification could be found for (A.2), which incorporates the concepts of continuous metric space and looks quite inappropriate for the discrete lattice. In spite of almost the same shape of the distributions (Figure 2), the variables (A.2) lead to confusion in interpretation of temperature that was a measure of the mean occupation number or mean energy for Planck's oscillators (4) and Wien's distribution (A.4).

Wilhelm Wien introduced his formula in 1896. It is quite close to Planck's radiation law and was one of the best candidates to describe experimental blackbody radiation spectrum. Friedrich Paschen in 1897 came to this conclusion about Wien's formula: *'It would seem very difficult to find another function that represents data with as few constants.'* (quoted from [11]).

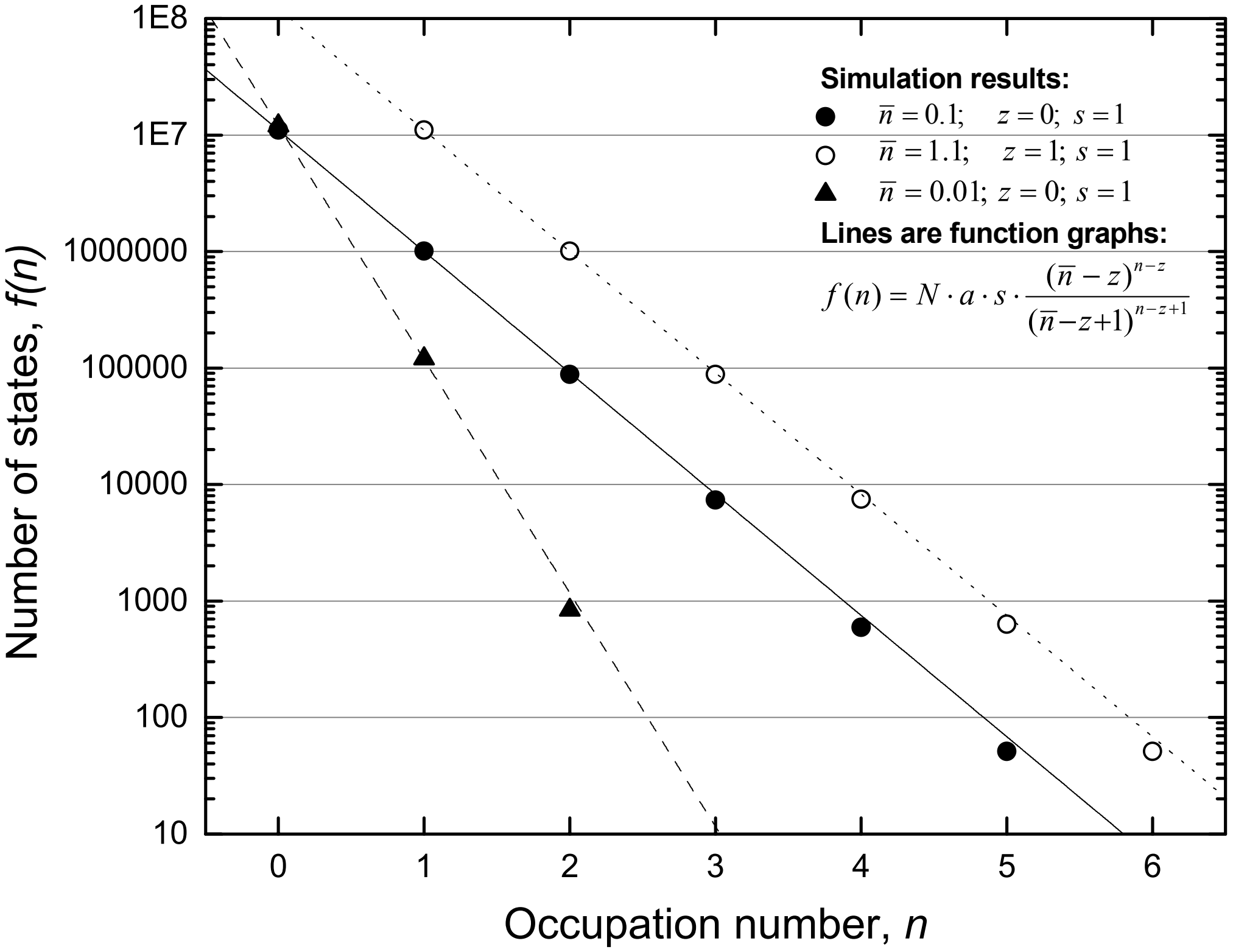

**FIGURE 1: Occupation Number Distribution of IHPP Gas Cell Attributes.**

IHPP gas exhibits an exponential occupation number distribution with inclination – ( $\bar{n} - z$ ). It is shifted by $z$ along axis *n*.

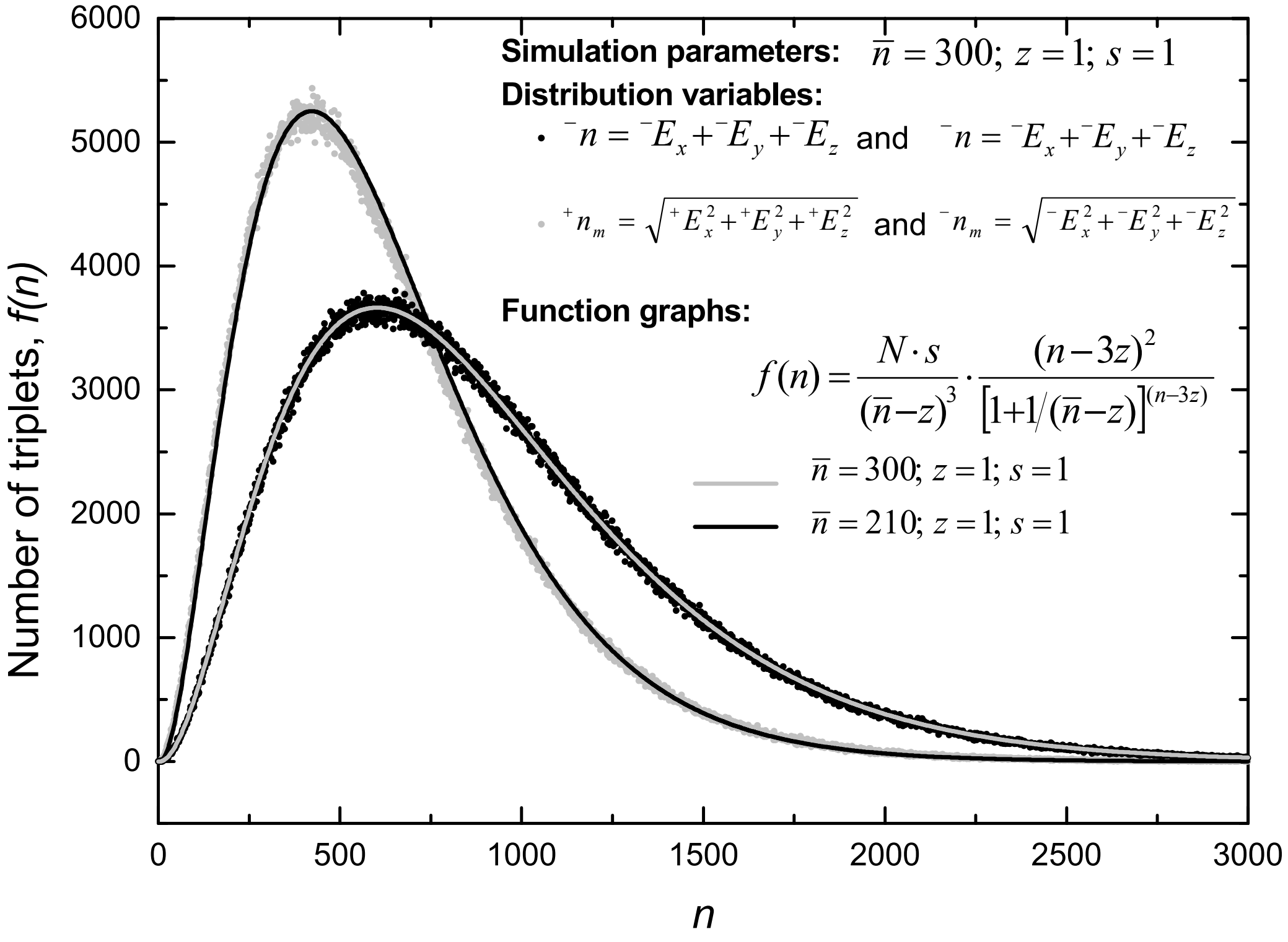

**FIGURE 2**

Distribution variables are constructed as triplets from all six characteristics of each cell.